\documentclass[a4paper]{article}

\usepackage{INTERSPEECH2020}
\usepackage[table,xcdraw]{xcolor}

\title{DNN No-Reference PSTN Speech Quality Prediction}
\name{Gabriel Mittag$^1$, Ross Cutler$^2$, Yasaman Hosseinkashi$^2$, Michael Revow$^2$, Sriram Srinivasan$^2$, Naglakshmi Chande$^2$, Robert Aichner$^2$}
\address{
  $^1$TU Berlin, $^2$Microsoft Corp.}
\email{gabriel.mittag@tu-berlin.de, first.last@microsoft.com}

\begin{document}
\maketitle
\begin{abstract}
Classic public switched telephone networks (PSTN) are often a black box for VoIP network providers, as they have no access to performance indicators, such as delay or packet loss. Only the degraded output speech signal can be used to monitor the speech quality of these networks. However, the current state-of-the-art speech quality models are not reliable enough to be used for live monitoring. One of the reasons for this is that PSTN distortions can be unique depending on the provider and country, which makes it difficult to train a model that generalizes well for different PSTN networks. In this paper, we present a new open-source PSTN speech quality test set with over 1000 crowdsourced real phone calls. Our proposed no-reference model outperforms the full-reference POLQA and no-reference P.563 on the validation and test set. Further, we analyzed the influence of file cropping on the perceived speech quality and the influence of the number of ratings and training size on the model accuracy. 
\end{abstract}
\section{Introduction}
Speech communication usage has been shifted in the past, firstly from landline to mobile and then to VoIP (Voice over IP) networks, where service users often use over-the-top providers. However, the classic public switched telephone network (PSTN) is still widely used and makes up a large portion of phone calls. To monitor the speech quality of a communication network, providers use instrumental speech quality prediction models. The ground truth of these models is the speech quality score derived from auditory listening tests conducted according to ITU-T Rec. P.800 \cite{p800}. In these experiments, naïve test participants listen to prerecorded speech samples and score them on a 5-point absolute category scale. The average across all participants then gives the so-called MOS (Mean Opinion Score). Recently, these listening experiments have been conducted through crowdsourcing rather than in the lab. The new ITU-T Recommendation P.808 \cite{P808} specifies how to conduct subjective speech quality experiments in the crowd with services such as Amazon Mechanical Turk (AMT) and how to screen the obtained data for unreliable ratings.

To monitor their networks, most providers use network parameters, such as packet loss and delay, to estimate the quality. An alternative is to send a reference speech probe through the channel. The degraded output speech signal can then be compared with the clean reference to estimate the speech quality with models such as PESQ \cite{p862, P862.2}, or the current ITU-T Recommendation for speech quality prediction POLQA \cite{P863}. The drawback of this method is that only a small subset of connections between fixed endpoints can be monitored. Parametric models, on the other hand, can be used for passive live monitoring of networks. However, if a phone call is made via multiple networks, for example, a call from a PSTN to a packet-switched VoIP network, no network performance indicators are available for a segment of the communication channel. The PSTN network is a black box for the VoIP service provider. In this case, only the degraded output speech signal is available, and thus POLQA or other full-reference speech quality models cannot be applied. However, in particular, the quality of PSTN networks can be challenging to estimate because the calls are often routed through multiple network providers, potentially in different countries. These circumstances lead to numerous transcoding with different codecs and can introduce obscure distortions that may only occur for a particular connection. 

The current state-of-the-art for no-reference speech quality prediction is ITU-T Rec. P.563 \cite{P563}, which is known for being unreliable for packet loss or live talking conditions \cite{Hines2015MeasuringAM}. Because of this recently, many no-reference speech quality models based on deep learning have been proposed \cite{Soni2016, Fu2018, Avila2019, mittag_a, Gamper, Ooster2019, Cauchi2019, Catellier2020, Dong20}. In this paper, we propose a new model that is trained and evaluated on real PSTN call recordings from 80 different providers in more than 50 countries. For a test set, we crowdsourced over 1000 real “PSTN to VoIP” calls and annotated each file in AMT with 30 ratings using P.808. This new real-call test set is made publicly available. Furthermore, we analyze the trade-off between the number of MOS ratings vs. training size on the model accuracy and show that the proposed model outperforms P.563 significantly on the open-sourced test set.
\section{Dataset}
\subsection{Training/Validation dataset design}
\begin{figure}[!ht]
    \begin{center}
        \includegraphics[width=\linewidth]{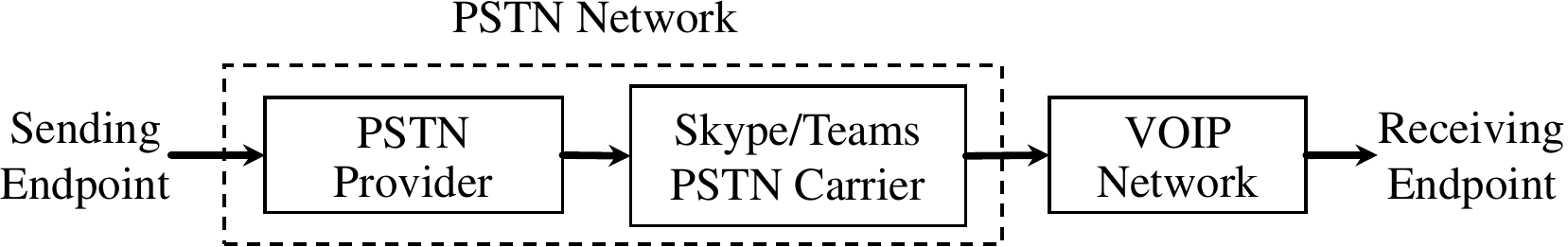}
    \end{center}
    \vspace{-0.4cm}
    \caption{Training set generation with automated phone calls}
    \label{fig:train_set}
\end{figure}

%
The training and validation data set was created by automatically conducting phone calls between a PSTN and a VoIP endpoint. As outlined in Figure \ref{fig:train_set}, the speech clips were sent from a sending endpoint through 80 different PSTN provider networks in over 50 countries. After going through a gateway, the signals were routed across one of five PSTN carrier networks that Skype and Microsoft Teams are using. Then, the signal was finally routed to the Skype VoIP network, where the degraded signal was recorded at the receiving endpoint.

The clean reference files used for the phone calls are derived from the public audiobook dataset Librivox\footnote{https://librivox.org}. The Librivox corpus contains recordings of 11,350 volunteers reading public domain audiobooks. Because many of the recordings are of poorer quality, the files have been filtered according to their quality as described in \cite{DNSChallenge2020}, leaving in a total 441 hours from 2150 speakers of good quality speech. These audiobook chapters were then segmented into 10 seconds clips and filtered for having a speech activity of at least 50\% (according to ITU-T Rec. P.56 \cite{P56}).
Since, in practice, there are often environmental sounds present during phone calls, we used the DNS-Challenge \cite{DNSChallenge2020} repo\footnote{https://github.com/microsoft/DNS-Challenge} to add background noise. The noise clips are taken from Audioset \cite{audioset}, Freesound, and the DEMAND \cite{demand} corpus and added to the clean files with an SNR between 0-40 dB. The resulting clean and noisy clips were played back by the PSTN bot that called the SfB bot. The degraded speech files on the receiving SfB bot side was then recorded. 

Overall, we conducted more than half a million automated phone calls. Because most of these calls were of good quality, we sampled a subset by putting less weight on files with a high POLQA MOS score, while maintaining clip and provider diversity. As a result, 79,980 degraded speech clips with a duration of 10 seconds were collected, 49,984 files based on noisy reference files, and 29,996 files based on clean reference files. The files were then split into a training and validation set. To make sure there is no speaker overlap between both sets, we used the speech files of 250 randomly selected speakers as a validation set, yielding a training size of 66,635 files and validation size of 13,345 files.

\subsection{Test set design}
Although we created a large collection of phone calls with a wide variety of different speakers and carriers, the phone calls are still simulated and do not contain the terminal device or calls from mobile networks. To have an even more realistic test set to evaluate the trained model, we asked crowdworkers on AMT to call a SkypeIn mailbox from their landline or mobile phone.  The crowdworkers then left a voicemail by reading out 10 random sentences taken from Graz University’s pitch-tracking dataset \cite{graz}. Overall we collected 1040 voicemail recordings (95\% from the U.S., the rest from India, Canada, U.K, and others). We randomly picked a 10-second clip for each voicemail recording that had more than 50\% speech activity.
\subsection{Listening experiment (P.808)}
The perceived speech quality of the training and test set were annotated in a listening experiment on AMT, according to P.808 \cite{P808}. Each training set file was rated by 5 participants, while the test set files were rated by 30 participants to ensure a low confidence interval of the MOS values for the model evaluation. The participants of all experiments were presented with the same six training files that cover the full quality range. Before calculating the MOS values, the ratings were screened against outliers and unexpected behavior from the crowdworkers. As a consequence, the resulting number of ratings of an individual file may be less than 5 for the training set or 30 for the test set, depending on the screening.
\subsection{Distribution of training and test sets}
\begin{figure}[!ht]
    \begin{center}
        \includegraphics[width=\linewidth]{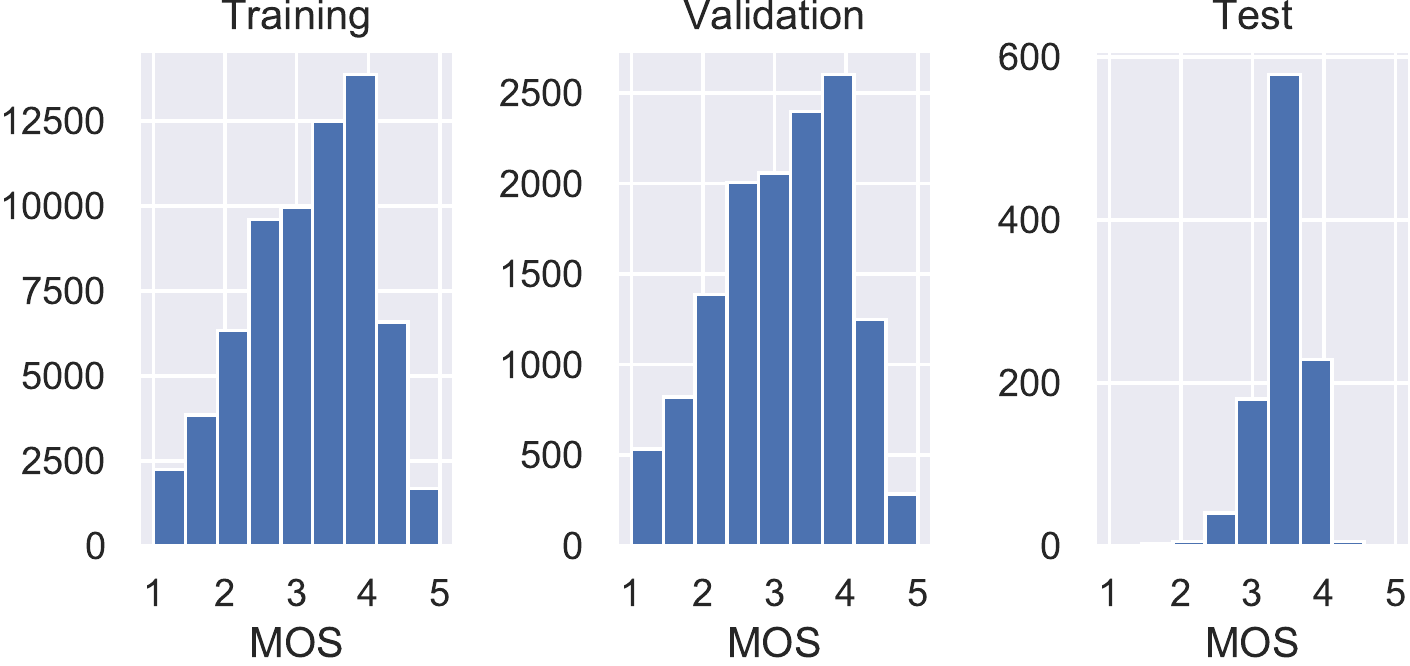}
    \end{center}
    \vspace{-0.4cm}
    \caption{Datasets speech quality histograms}
    \label{fig:hist}
\end{figure}

The quality distributions of the datasets are presented in Figure \ref{fig:hist}. It can be seen that most files were rated with a MOS score around 3-4, with fewer files of poor or excellent speech quality. The distribution of the test set is highly skewed towards clips with ratings between 3.5 - 4. However, the test set represents a natural quality distribution of phone calls executed by crowdworkers. Table \ref{tab:datasets} shows an overview of the datasets with the number of individual clips/sentences and speakers of the reference files. 
\begin{table}[htb]
\caption{Datasets}
\vspace{-0.3cm}
\centering
\label{tab:datasets}
\resizebox{0.95\linewidth}{!}{%
\begin{tabular}{l|rrrrr}
      & \multicolumn{1}{c}{\# Files} & \multicolumn{1}{c}{\# Clean} & \multicolumn{1}{c}{\# Noisy} & \multicolumn{1}{c}{\# Sentences} & \multicolumn{1}{c}{\# Speakers} \\ \hline
Train & 66,635                       & 25,896                       & 40,739                       & 24,316                           & 1,454                           \\
Val   & 13,345                       & 4,100                        & 9,245                        & 6,866                            & 250                             \\
Test  & 1,040                        & Real Call                         & Real Call                         & 1,040                            & 1,040                          
\end{tabular}
}
\end{table}
\subsection{Opensource link (TBD)}
Only few speech datasets with subjective quality ratings are publicly available. ITU-T P Suppl. 23 \cite{supp23} contains noise, codec, and packet loss conditions, the NOIZEUS \cite{NOIZEUS} dataset was designed to evaluated noisy speech that is processed with speech enhancement algorithms, and the TCD-VoIP dataset \cite{Harte} contains speech processed with various VoIP conditions. All of these datasets contain only simulated conditions with pre-recorded, typical P.800 double sentences. However, such speech samples are not realistic in a no-reference scenario, where the prediction model is applied to live phone calls. 

To overcome this lack of realistic datasets, we are open sourcing the test set, comprising of 1040 clips extracted from live phone calls, under following link: TBD
\section{Model description}
\begin{figure}[!ht]
    \begin{center}
        \includegraphics[width=\linewidth]{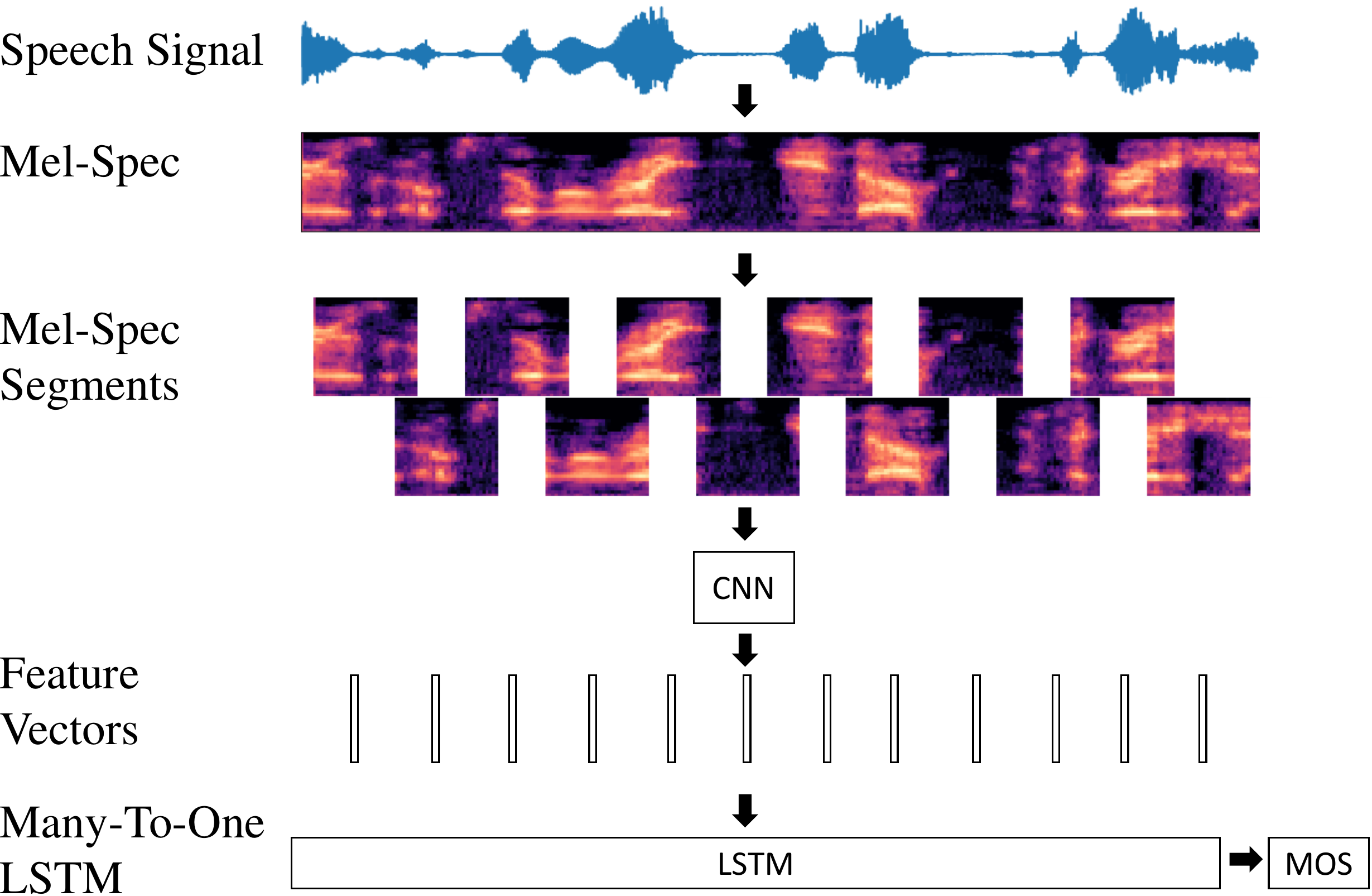}
    \end{center}
    \vspace{-0.4cm}
    \caption{PSTN speech quality model}
    \label{fig:model}
\end{figure}

The speech quality prediction model used in this paper is a narrowband (up to 4 kHz) version of the CNN-LSTM neural network presented in \cite{mittag_a, mittag_b}\footnote{https://github.com/gabrielmittag/NISQA}. Instead of using mel-spectrograms with a maximum frequency of 16 kHz and 48 mel bands, the mel spectrogram inputs to the proposed PSTN model have a maximum frequency of 4 kHz and 32 mel bands. Figure \ref{fig:model} shows the design of the proposed PSTN speech quality model. At first, mel-spectrograms with a FFT window length of 20 ms and a hop size of 10 ms are calculated from the speech signal. These mel-spectrograms are then divided into segments with a width of 33 (i.e. 330 ms) and a height of 32 (corresponding to the 32 mel bands). The hop size between the segments is 24, which leads to a segment overlap of 27\% and overall 41 segments for a 10 sec speech signal. Each segment is then processed by the same CNN network (see Table \ref{tab:cnn}), which outputs a CNN feature vector of length 10. This 41x10 feature sequence is used as input to a bidirectional LSTM network with 2x50 hidden units. The output of the last LSTM time step is then finally used to estimate the overall MOS score.
\begin{table}[htb]
\caption{CNN design (each convolutional layer is followed by a batch normalization and ReLu layer. The kernel size is 3x3.)}
\vspace{-0.3cm}
\centering
\label{tab:cnn}
\resizebox{0.65\linewidth}{!}{%
\begin{tabular}{l|l}
\textbf{Layer}                  & \textbf{Output size} \\ \hline
Input                  & 41x1x32x33  \\ \hline
Conv 1                 & 41x16x32x33 \\
Pool                   & 41x16x16x16 \\ \hline
Conv 2                 & 41x16x16x16 \\
Pool / Dropout(20\%)   & 41x16x8x8   \\ \hline
Conv 3                 & 41x32x8x8   \\
Conv 4                 & 41x32x8x8   \\
Pool / Dropout(20\%)   & 41x32x4x4   \\ \hline
Conv 5 / Dropout(20\%) & 41x32x4x4   \\
Conv 6 (no padding)    & 41x32x1x1   \\ \hline
FC                     & 41x10      
\end{tabular}
}
\end{table}
\section{Experiments}
\subsection{Influence of file cropping on speech quality perception}
\begin{figure}[!ht]
    \begin{center}
        \includegraphics[width=\linewidth]{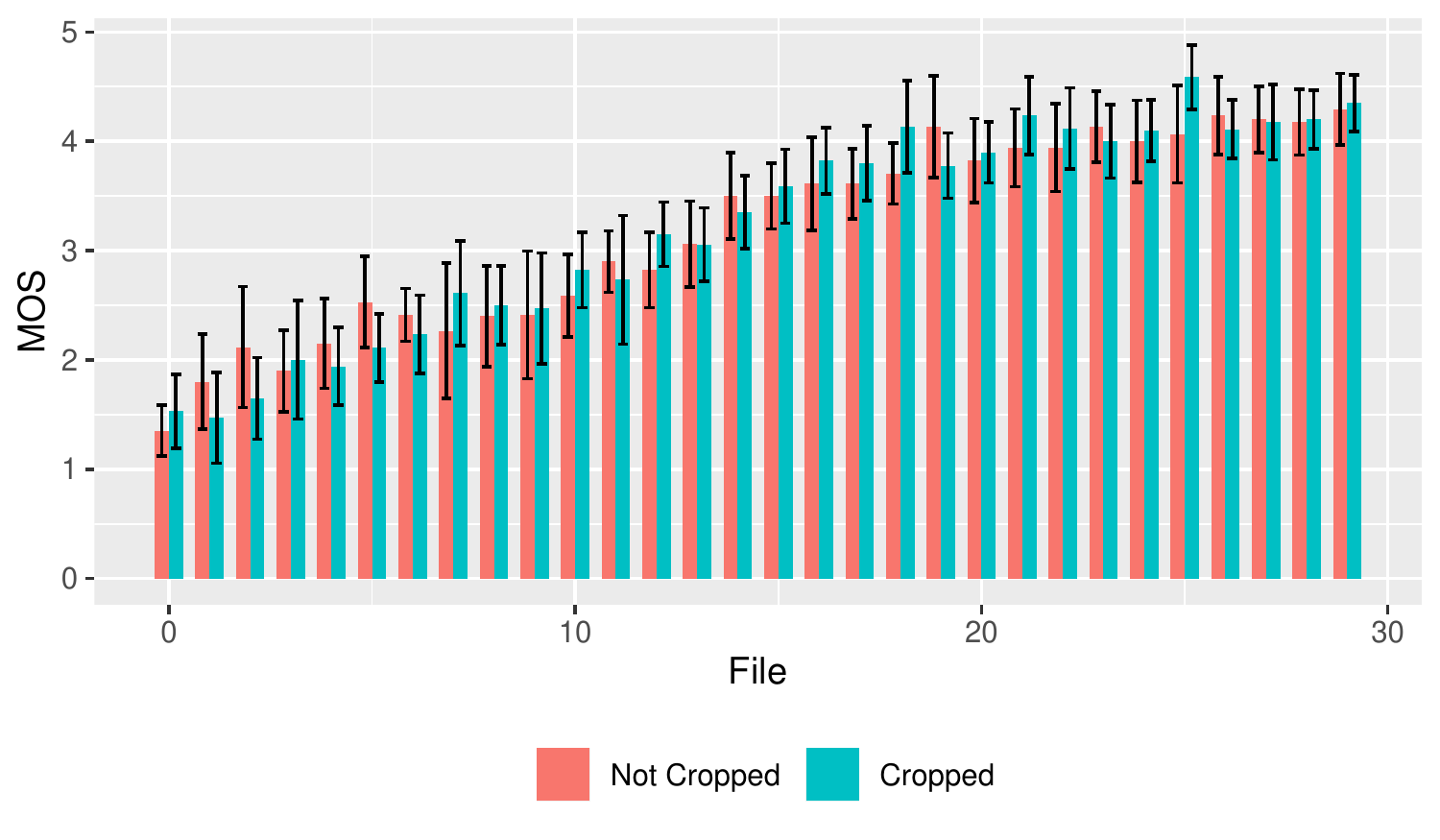}
    \end{center}
    \vspace{-0.4cm}
    \caption{Cropped vs. not Cropped}
    \label{fig:crop}
\end{figure}

Because we segmented the audiobooks and the live calls into clips with 10 seconds duration, we wanted to know if cropping a file in the middle of a word influences the perceived speech quality. One common distortion in speech communication networks are lost packets that lead to interruptions. Therefore, participants might judge a cropped file with an interruption at the start or the end of the signal as distorted by packet loss. To analyze the influence, we conducted two P.808 listening experiments with each 20 different participants. In the first experiment, half of the files were manually cut during silent segments of the speech signal and therefore contained no interruptions, the other half was cropped during an active speech segment of the file. In the second experiment, the files that were previously cropped with an interruption were included without interruptions and vice versa. As a result, each experiment contained 30 files (15 files cropped/with interruptions and 15 not cropped/without interruptions). Figure \ref{fig:crop} shows the MOS scores of all 60 files, where a cropped file is grouped with its non-cropped version. It can be seen that the cropping does not influence the speech quality, as the confidence intervals are clearly overlapping. Furthermore, a paired t-test revealed no significant difference between the cropped and non-cropped files (p-value=0.49).
\subsection{Number of ratings vs model accuracy}
Before we labeled the overall 81,020 files of the datasets in P.808 listening experiments, we analyzed how many ratings per file are necessary for the model training. To this end, we labeled a pilot dataset with 6500 clean training files and 1203 clean validation files and 10 ratings per file. Then we ran the training for 1-8 subjective MOS ratings and evaluated the model accuracy in terms of Pearson's correlation coefficient (PCC). The training set for this experiment is a 5120 files subset of the clean training data for which at least 8 votes were available after data cleansing. Before training the files, the MOS scores were recalculated with a random selection of $x$ number of ratings. The random selection of ratings, recalculation of the MOS scores, and model training was then repeated 7 times. The results on the clean validation subset are presented on the left-hand side of Figure \ref{fig:train_votes}. It can be seen that the model accuracy increases most rapidly for up to 3 ratings. However, the performance increase is still not saturated for 8 ratings. It should also be noted that the model performance will depend on the random selection of ratings and, therefore might be, on average lower for a lower number of ratings.
%
%
\subsection{Training size influence}
\begin{figure}[!ht]
    \begin{minipage}{0.48\linewidth}
        \begin{center}
            \includegraphics[width=\linewidth]{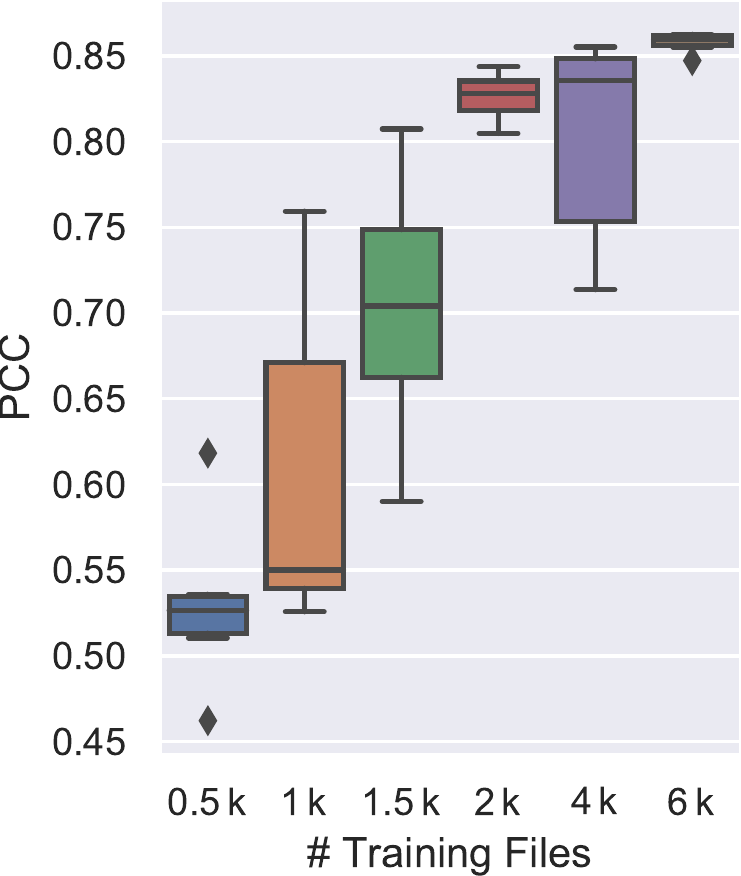}
        \end{center}
    \end{minipage}
    \hfill
    \begin{minipage}{0.48\linewidth}
        \begin{center}
            \includegraphics[width=\linewidth]{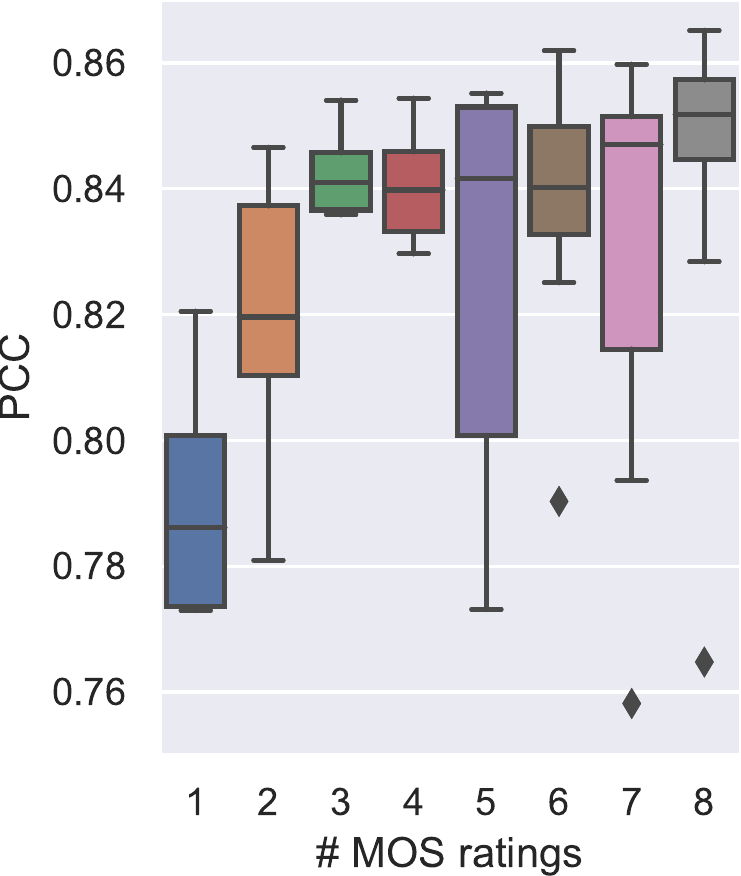}
        \end{center}
    \end{minipage} 
    \vspace{-0.2cm}
    \caption{Influence of training size and number of ratings on model accuracy}
    \label{fig:train_votes}
\end{figure}

Apart from the number of ratings, we also analyzed the training size influence on the model accuracy. For this purpose, we randomly subsampled files from the pilot dataset and trained the model six times for each training size. The results are presented on the right-hand side of Figure \ref{fig:train_votes} and show that there is a sharp performance increase for up to 2000 training files. However, for training sizes larger than 2000, the model accuracy is still further improving.
%
%

\subsection{Training size vs. number of ratings}
\begin{figure}[!ht]
    \begin{center}
        \includegraphics[width=0.5\linewidth]{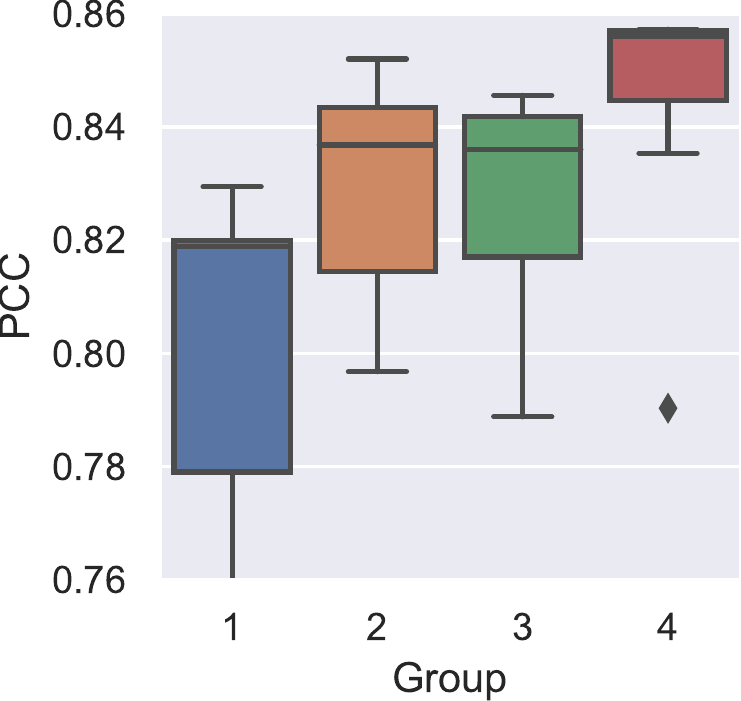}
    \end{center}
    
    \begin{center}
        \resizebox{0.6\linewidth}{!}{%
        \begin{tabular}{|
        >{\columncolor[HTML]{FFFFFF}}l |
        >{\columncolor[HTML]{ABC6F0}}c 
        >{\columncolor[HTML]{F9CBB2}}c 
        >{\columncolor[HTML]{ADD6B6}}c 
        >{\columncolor[HTML]{E7A8AA}}c |}
        \hline
        {\color[HTML]{000000} \textbf{Group}} & {\color[HTML]{000000} \textbf{1}} & {\color[HTML]{000000} \textbf{2}} & {\color[HTML]{000000} \textbf{3}} & {\color[HTML]{000000} \textbf{4}} \\ \hline
        {\color[HTML]{000000} Files}  & {\color[HTML]{000000} 2500}       & {\color[HTML]{000000} 5000}       & {\color[HTML]{000000} 2500}       & {\color[HTML]{000000} 5000}       \\
        {\color[HTML]{000000} Ratings}          & {\color[HTML]{000000} 4}          & {\color[HTML]{000000} 4}          & {\color[HTML]{000000} 8}          & {\color[HTML]{000000} 8}          \\ \hline
        \end{tabular}
        }    
    \end{center}
    \vspace{-0.4cm}
    \caption{Boxplot of 7 training runs of training size vs. number of votes analysis}
    \label{fig:votes_vs_tr_size}
\end{figure}

Lastly, we analyzed whether it is better to add more ratings or more files to the training set to improve the model accuracy. We created four different groups as shown in Figure \ref{fig:votes_vs_tr_size}. The first training set group has 2500 files with each 4 votes, Group 2 has the same number of ratings, but twice as many training files, Group 3 has twice as many ratings as Group 1 but the same number of training files, Group 4 has twice as many ratings and twice as many training files. We ran the model training 6 times for each group, the results can be seen in Figure \ref{fig:votes_vs_tr_size}. The boxplots show that the best model performance was, as expected, achieved by Group 4. But still, the model performance also improved significantly for both Group 2 and 3. In direct comparison, the experiments showed that it is better to add more files rather than more MOS ratings, as Group 2 outperforms Group 3. These results cannot necessarily be generalized as the outcome may change depending on how many files and ratings are analyzed in particular. However, they indicate that more diversity in terms of speakers, sentences, and quality distortions are more helpful than a higher MOS accuracy for neural network training. Based on these results, we chose a larger training size with over 80,000 files in total and labeled the speech files for the training and validation corpus with 5 instead of 10 ratings each.
%
%
\subsection{Model results}
\begin{figure}[!ht]
      \begin{minipage}{0.48\linewidth}
            \begin{center}
            \centerline{PSTN Model}\medskip
            \includegraphics[width=\linewidth]{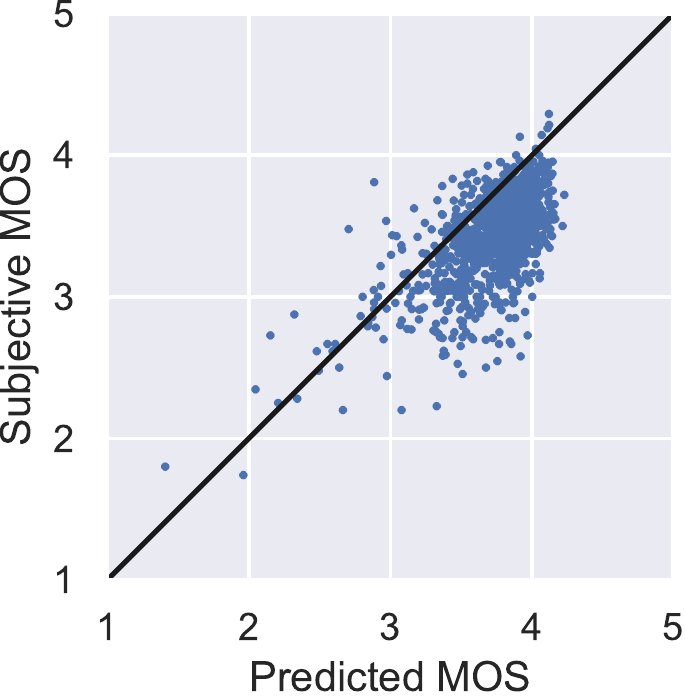}
          
          \end{center}
      \end{minipage}
        \hfill
      \begin{minipage}{0.48\linewidth}
       \centerline{P.563}\medskip
            \includegraphics[width=\linewidth]{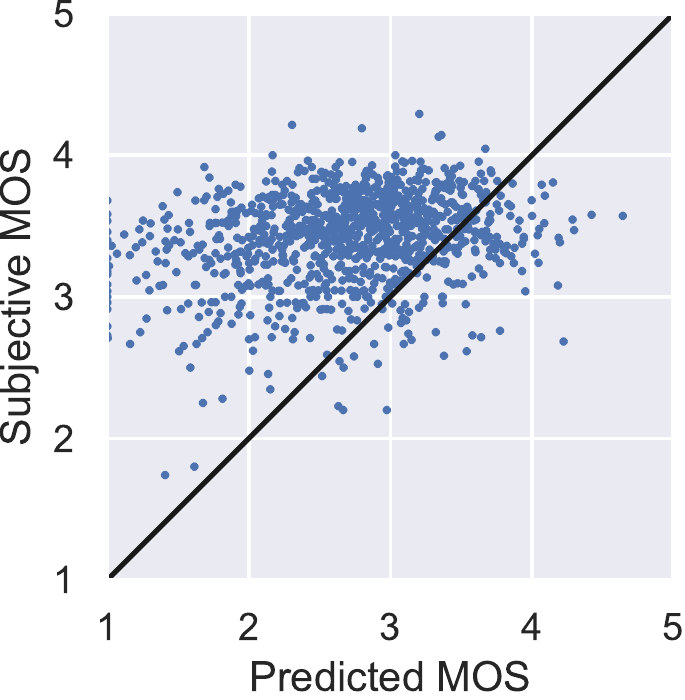}
      \end{minipage} 
\caption{Test set correlation diagrams}
\label{fig:results_test}
\end{figure}

The model was trained with a learning rate of 0.001, mini-batch size of 200 and Adam optimizer with mean-square error loss. The best resulting model performance after multiple training runs is presented in Table \ref{tab:results} in terms of root-mean-square error (RMSE) and PCC. The table shows that the model outperforms the full-reference model POLQA on the clean validation set with a PCC of 0.87 compared to 0.78. The overall PCC on the validation set is 0.80, where the correlation for the noisy data is notably lower with a PCC of 0.76. The PCC on the test set is relatively low, with only 0.63. However, this test set is challenging to predict since it consists of live phone calls, which were not contained in the training data. Furthermore, the distribution of the test set is highly skewed, as can also be seen in the correlation diagram in Figure \ref{fig:results_test}. Most of the files have scores that are very close to each other with MOS values around 3.5-4, which make a high correlation more difficult. The proposed model still significantly outperforms the current ITU-T recommended model for no-reference narrowband speech quality prediction P.563, which only achieves a PCC of 0.25.
\begin{table}[htb]
\caption{Model results in terms of root-mean-square error and Pearson's correlation}
\vspace{-0.3cm}
\centering
\label{tab:results}
\resizebox{0.95\linewidth}{!}{%
\begin{tabular}{l|cc|cc|cc}
          & \multicolumn{2}{c|}{Proposed} & \multicolumn{2}{c|}{POLQA} & \multicolumn{2}{c}{P.563} \\
Data Set  & PCC            & RMSE           & PCC          & RMSE        & PCC         & RMSE        \\ \hline
Val Clean & 0.87           & 0.50           & 0.78         & 0.72        & 0.49        & 0.83        \\
Val Noisy & 0.76           & 0.55           &              &             &             &             \\
Val Total & 0.80           & 0.53           &              &             &             &             \\
Test      & 0.63           & 0.41           &              &             & 0.25        & 0.97       
\end{tabular}
}
\end{table}
To see how the model performs on a more evenly distributed dataset, we randomly sampled uniform subsets, where the files were divided into 3 bins according to their MOS values ([1.5-2.5), [2.5-3.5), [3.5-4.5]) and drew 13 files from each bin, resulting in 39 files per subset. We repeated this process of random sampling 1000 times and calculated the average PCC. In Table \ref{tab:results_2} it can be seen that results on the clean validation data for the proposed PSTN model and POLQA did not change when compared to the original data set in Table \ref{tab:results}. The results on the skewed test set, however, increased to 0.84, which is higher than the PCC POLQA achieved on the clean validation set. 
\begin{table}[htb]
\caption{Average results after repeatedly sampling uniform subset 1000 times. Note POLQA is full-reference.}
\vspace{-0.3cm}
\centering
\label{tab:results_2}
\resizebox{0.95\linewidth}{!}{%
\begin{tabular}{l|cccc}
               & \multicolumn{4}{c}{PCC}                    \\ \hline
Model          & Proposed     & P.563    & Proposed      & POLQA     \\
Data set       & Test Set & Test Set & Val Clean & Val Clean \\ \hline
Uniform subset & 0.84     & 0.49     & 0.87      & 0.78     
\end{tabular}
}
\end{table}
\section{Conclusions}
In this paper, we presented a PSTN speech quality model that significantly outperforms the current state-of-the-art speech quality prediction model P.563 and also shows to perform better than the full-reference model POLQA. We tested the model on a new PSTN speech quality dataset with crowdsourced PSTN to Skype mailbox calls that we annotated according to ITU-T Rec. P.808. Additionally, we analyzed the influence of cropping on the perceived speech quality. We also showed that only 3 MOS ratings per file can be enough to train a neural network speech quality prediction model if enough training files are available. Since there has been a lack of realistic datasets in the speech quality field, we are open sourcing our dataset so researchers have a common test set for future work.
\bibliographystyle{IEEEtran}
\bibliography{main}
\end{document}